\documentclass[9pt,conference]{IEEEtran}
\usepackage{dcase2025}

\usepackage{bm} 

\usepackage{dcase2025,amsmath,graphicx,url,times,booktabs, tabularx}
\usepackage{graphicx}
\usepackage{soul,color}
\usepackage{caption}
\usepackage{subcaption}
\usepackage{amssymb}
\usepackage{parskip}
\usepackage{setspace}
\usepackage{hyperref}

\def\etal{\emph{et al}.\ }

\title{Integrating Spatial and Semantic Embeddings for Stereo Sound Event Localization in Videos}

\twoauthors
 {Davide Berghi\sthanks{This research was funded by EPSRC-BBC Prosperity Partnership `{AI4ME}: Future personalised object-based media experiences delivered at scale anywhere' (EP/V038087/1). For the purpose of open access, the authors have applied a Creative Commons Attribution (CC BY) license to any Author Accepted Manuscript version arising. Data supporting this study is available from \url{https://zenodo.org/records/15559774}}}
   {University of Surrey, CVSSP\\
    Guildford, UK \\
    davide.berghi@surrey.ac.uk}
 {Philip J. B. Jackson}
   {University of Surrey, CVSSP\\
    Guildford, UK \\
    p.jackson@surrey.ac.uk}

\begin{document}

\maketitle

\begin{abstract}

In this study, we address the multimodal task of stereo sound event localization and detection with source distance estimation (3D SELD) in regular video content.
3D SELD is a complex task that combines temporal event classification with spatial localization, requiring reasoning across spatial, temporal, and semantic dimensions. The last is arguably the most challenging to model. Traditional SELD approaches typically rely on multichannel input, limiting their capacity to benefit from large-scale pre-training due to data constraints.
To overcome this, we enhance a standard SELD architecture with semantic information by integrating pre-trained, contrastive language-aligned models: CLAP for audio and OWL-ViT for visual inputs. These embeddings are incorporated into a modified Conformer module tailored for multimodal fusion, which we refer to as the Cross-Modal Conformer.
We perform an ablation study on the development set of the DCASE2025 Task3 Stereo SELD Dataset to assess the individual contributions of the language-aligned models and benchmark against the DCASE Task 3 baseline systems.
Additionally, we detail the curation process of large synthetic audio and audio-visual datasets used for model pre-training. These datasets were further expanded through left-right channel swapping augmentation.
Our approach, combining extensive pre-training, model ensembling, and visual post-processing, achieved second rank in the DCASE 2025 Challenge Task 3 (Track B), underscoring the effectiveness of our method. Future work will explore the modality-specific contributions and architectural refinements.

\end{abstract}


\section{Introduction}
\label{sec:intro}

Sound Event Localization and Detection \cite{Adavanne:2019:SELDnet} is a combined task that integrates sound event detection (SED) \cite{Adavanne:2017:sed} and sound source localization (SSL) \cite{Adavanne:2018:DOA}. The goal is to identify active sound events from predefined target classes, track their temporal activity, and estimate their spatial positions.
SELD systems are crucial for a wide range of real-world applications, including human-robot interaction \cite{Xinyuan:2023:AVcrossAtt}, security monitoring, and immersive media production \cite{Berghi:2024:forecasterFlexOBM}. 

Since 2019, SELD has been a dedicated task in the Detection and Classification of Acoustic Scenes and Events (DCASE) Challenge.
Over successive editions, the task has evolved to include increasingly complex scenarios, such as moving sound sources \cite{politis:2020:DCASE}, ignoring external interfering sounds \cite{politis:2021:DCASE}, incorporating visual input to enable multimodal SELD in 360$^{\circ}$ videos \cite{Shimada2023STARSS23AA}, and estimating source distance \cite{Diaz-Guerra:2024:seldBaseline24}.
For the 2025 edition, the challenge has shifted from the traditional 4-channel first-order ambisonics (FOA) and microphone array (MIC) formats to stereo SELD using conventional frontal video content, i.e., leveraging only left and right audio channels and perspective video. This format is better aligned with the requirements of conventional media content.
In stereo SELD the DOA prediction is limited to the azimuth angles in the range [-90$^{\circ}$, 90$^{\circ}$], as elevation 
is ill-defined when using
just two horizontally arranged channels, and distinguishing between front and back sources is inherently ambiguous. Source distance estimation, introduced in the previous edition, remains part of the task, which is therefore often referred to as 3D SELD. Additionally, in the audio-visual track, a new subtask has been introduced: predicting whether sound sources are onscreen or offscreen.

SELD is a non-trivial task, as it requires reasoning across spatial, temporal, and semantic dimensions: spatial information gives cues for direction and distance estimation;
temporal information marks movements and onsets/offsets of sound source activity; semantic information identifies objects, their relations and likely behaviors.
Recent advances in large language models (LLMs) \cite{Brown:2020:GPT}, vision-language models (VLMs) \cite{Radford:2021:clip}, and audio-language models (ALMs) \cite{Huang:2024:AudioGPT} demonstrate that language is a powerful lens for enabling semantic understanding and complex reasoning in relation to media content. 
Building on this, we posit that leveraging language-aligned models can enhance a SELD model’s semantic reasoning and hence indirectly benefit spatial and temporal reasoning. 
Traditionally, spatial localization relies on multichannel audio, allowing models to infer source positions through inter-channel time and level differences. This dependence limits the use of large-scale pre-training datasets. 
While synthetic audio \cite{Roman:2024:spatialScaper} and visual \cite{roman2025generating} data can partially address this limitation, effective language alignment typically demands extensive pre-training on large-scale, heterogeneous data.
Therefore, we extend our SELD architecture by integrating two existing language-aligned models: CLAP \cite{wu:2023:clap_laion} for the audio modality and OWL-ViT \cite{Minderer:2022:owl_vit} for the visual modality. Specifically, we extract audio and visual embeddings using their respective pretrained encoders and combine them with SELD embeddings obtained from a CNN-Conformer backbone, a widely adopted architecture in SELD research \cite{Wang:2023:ACS,Berghi:2024:ICASSP24,Xue:2023:resnetConf}. We argue that the embeddings from CLAP and OWL-ViT are semantically rich and provide complementary information to the task.
To fuse these embeddings, we employ an adapted version of the Conformer architecture \cite{Gulati2020ConformerCT}, which we refer to as the Cross-Modal Conformer (CMC)\@. This module is used to integrate intra-modal embeddings from different sources (e.g., the SELD audio encoder and CLAP) and inter-modal embeddings (i.e., the combined audio representation and the visual embedding from OWL-ViT).
This architecture is outlined in Sec.~\ref{subsec:model}, and an ablation study to demonstrate the individual contributions of the language-aligned
models and benchmark against the DCASE Task 3 baseline systems is presented in Sec.~\ref{sec:exp}.


\begin{figure}[bt]
\centering
  \includegraphics[width=\columnwidth]{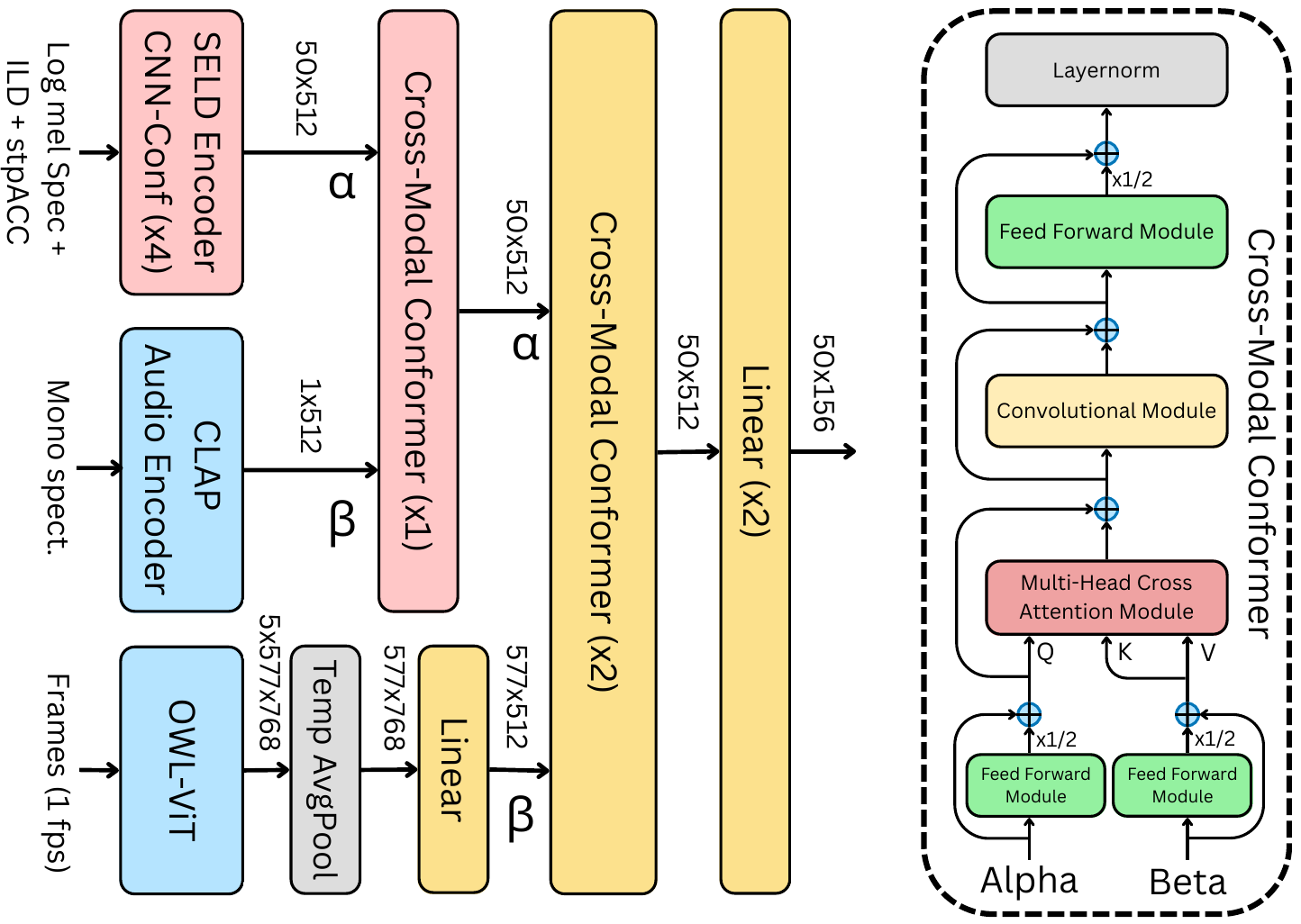} \\ [1ex]
\caption{Proposed audio-visual model (left). Blue blocks are frozen, red ones are pre-trained on the audio 5k and audio-visual 2k datasets, yellow on audio-visual 2k dataset only. 
Cross-Modal Conformer (CMC) architecture adapting the original Conformer \cite{Gulati2020ConformerCT} to process two generic modalities ``Alpha'' $\alpha$ and ``Beta'' $\beta$ (right).}
\label{fig:model} 
\end{figure}

To tackle the distance estimation subtask, we partnered common SELD input features with short-term power of the autocorrelation stpACC features \cite{Berghi:2025:distanceFeat} in Sec.~\ref{subsec:features}.  
Furthermore, we synthesized and carefully curated large audio and audio-visual datasets used to pre-train our models, described in Sec.~\ref{subsec:preproc}.
Final refinements includes a visual post-processing step based on human keypoint detection and model ensembling, which is described in Sec.~\ref{sec:exp} on experiments. Our best model ranked 2nd at the DCASE 2025 Challenge Task 3 (Track B), demonstrating the effectiveness of the proposed approach.

This paper introduces three key contributions: (i) the integration of semantically rich, language-aligned models into an audio-visual SELD architecture; (ii) an adapted Conformer architecture designed to fuse multiple modalities; (iii) substantial performance gains via curated data synthesis and engineering refinements, resulting in a second-place ranking in the DCASE2025 Task 3 Challenge (Track B).

\section{Method}
\label{sec:method}

\subsection{Proposed Architecture}
\label{subsec:model}

The proposed model includes a SELD encoder that extracts embeddings from multichannel inputs, which are fused with CLAP audio embeddings via a cross-modal cross-attention mechanism. We adapted the Conformer architecture to accommodate inputs from different modalities.
The resulting audio representation is then fused with visual features from OWL-ViT through a second Cross-Modal Conformer. A final feed-forward module predicts multi-ACCDDOA vectors for up to $N{=}3$ tracks \cite{Krause:2024:seldDistance}, including on/off-screen activity as in the challenge baseline. The model is trained using class-wise Auxiliary Duplicating Permutation Invariant Training (ADPIT) loss \cite{Shimada:2022:multiACCDOA,cao:2020:EIN,cao:2021:EINv2}.

\subsubsection{SELD Encoder}
\label{subsubsec:seld_enc}


We adopt a CNN-Conformer architecture for the SELD encoder, as it is widely adopted in SELD research \cite{Wang:2023:ACS,Berghi:2024:ICASSP24,Xue:2023:resnetConf}. The stereo input features of shape $C_\mathrm{in}{\times} T_\mathrm{in}{\times} F_\mathrm{in}$ are processed by four CNN blocks with residual connections, each comprising two 3$\times$3 convolutions, batch normalization \cite{ioffe:2015:BN}, ReLU, and stride-2 average pooling, halving the temporal and frequency dimension at each block. This reduces the temporal and frequency dimensions by a factor of 16, producing a $512{\times}T_\mathrm{in}/16{\times}F_\mathrm{in}/16$ tensor. After frequency pooling and reshaping, we obtain a $T_\mathrm{in}/16{\times}512$ embedding, aligned with the label frame rate of 10 labels\,/\,sec.. A four-layer Conformer with eight attention heads and depthwise convolutions (kernel size 51) \cite{Wang:2023:ACS} processes this sequence.

\subsubsection{Cross-Modal Conformer}
\label{subsubsec:cm_conf}

The CMC is an adapted version of the Conformer architecture proposed by Gulati \etal \cite{Gulati2020ConformerCT}. A schematic representation is shown on the left side of Fig.~\ref{fig:model}, where a generic modality ``Alpha'' ${\in}\,\mathbb{R}^{T_\alpha \times d_k}$ is combined with another modality ``Beta'' ${\in}\,\mathbb{R}^{T_\beta \times d_k}$ \cite{Xinyuan:2023:AVcrossAtt}.
Each sub-module retains the original Conformer structure, with two key changes: (i) two initial feed-forward layers independently process the modalities in parallel, and (ii) the standard multi-head self-attention is replaced by cross-attention, using queries from modality Alpha and keys/values from modality Beta. Our model includes two CMC blocks—first fusing SELD and CLAP embeddings, and then combining the result with OWL-ViT visual features. A single layer suffices for audio fusion, while two layers improve audio-visual fusion. Each CMC uses eight attention heads and depthwise convolutions with a kernel size of 51 \cite{Wang:2023:ACS}.

\subsubsection{Video Embedding Extraction}

Since SELD involves spatial, semantic, and temporal reasoning, our previous works \cite{Berghi:2024:ICASSP24,Berghi:2024:DCASE24techRep} employed ResNet50 \cite{He:2016:resnet} to extract visual features from each individual video frames, capturing the temporal dimension across time. To obtain per-frame visual embeddings, a $7{\times}7$ average pooling operation was applied, resulting in a single feature vector for each frame.
However, we now argue that this approach degrades spatial resolution, limiting the model’s ability to utilize fine-grained spatial cues. In contrast, other works, such as \cite{hong:2024:MVAnet} or this year’s baseline system, apply average pooling across the channel dimension of the ResNet50 output, preserving the $7{\times}7$ spatial layout. Yet, we believe that this alternative sacrifices semantic richness, as pooling across channels degrades the learned feature representations.

Effectively managing time, space, and channel information in a unified framework is non-trivial. Nevertheless, we argue that temporal dynamics can be effectively captured by the SELD encoder itself. As such, the visual processing branch should be optimized to better leverage semantic and spatial information.
To this end, we sacrifice temporal granularity in the visual stream. We replaced ResNet50 with OWL-ViT \cite{Minderer:2022:owl_vit}, a contrastive, language-aligned model, like CLAP, but specifically trained for visual grounding tasks like object detection. As a result, we expect OWL-ViT to produce visual embeddings that are both semantically and spatially rich.
To retain some temporal context without incurring high computational costs, we sample video frames at 1 fps. The resulting embeddings are then aggregated via temporal average pooling. All individual ViT token embeddings are maintained to preserve spatial and semantic information.

OWL-ViT requires square input frames of size $768{\times}768$. To avoid distorting the original rectangular frames in the dataset, we explored two pre-processing strategies during preliminary experiments: letterboxing the frames with black bars to preserve the aspect ratio, and applying a non-linear spatial transformation that preserves object proportions in the central region while progressively stretching the top and bottom areas. This second approach is based on the assumption that most sound events occur near the center of the frame.
The non-linear re-framing method yielded slightly better performance, and we therefore integrated it into our pre-processing pipeline.
Visual tokens are extracted from $32{\times}32$ pixel patches, resulting in 576 tokens per frame, plus an additional classification token. Each token embedding has a dimensionality of 768. To align with the model's architecture, we apply a linear projection to reduce this dimensionality to $d_k{=}512$. These OWL-ViT embeddings serve as key/value pairs (i.e., modality ``Beta'') in the second CMC.

\subsection{Acoustic Input Features}
\label{subsec:features}

Since the left and right audio channels in the dataset are arithmetically derived from FOA signals as described in \cite{Mazzon2019FirstOA} and \cite{Wilkins:2023:TwoVsFour,shimada:2025:SAVG}, rather than captured by two physically separated microphones, they should not present inter-channel time or phase differences. 
So, we adopted the inter-channel level difference (ILD) as the primary spatial feature for the SELD encoder, alongside log mel spectrograms computed independently from each channel. ILD features are calculated as the ratio of the squared magnitudes of the short-time Fourier transforms (STFTs) of the two channels, and subsequently mapped into the log mel domain: 
\begin{equation}
\mathbf{ILD}(m,t) = \log\left( \mathbf{H}_{\mathrm{mel}}  \frac{|\mathbf{L}(f,t)|^2+ \epsilon}{|\mathbf{R}(f,t)|^2 + \epsilon} \right),
\end{equation}
where $\mathbf{L}(f,t)$ and $\mathbf{R}(f,t)$ are the STFTs of the left and right channels, respectively, $\mathbf{H}_{\mathrm{mel}}$ is the mel filter bank, $\epsilon$ is a small constant to avoid division by zero and instability, and $m$ in the mel frequency index.
To further support the model in the distance estimation subtask, we include short-term power of the autocorrelation (stpACC) features, as proposed in \cite{Berghi:2025:distanceFeat}. 
The final concatenation of acoustic input features has a channel dimension of $C_\mathrm{in}{=}4$. Input features are normalized for zero mean and unit standard deviation vectors.

\subsection{Pre-processing and Data Augmentation}
\label{subsec:preproc}

We pre-trained our model on synthetic data while keeping the CLAP audio encoder and OWL-ViT weights frozen. Synthetic FOA audio was generated using SpatialScaper \cite{Roman:2024:spatialScaper}, which convolves FSD50K sounds \cite{fonseca2022FSD50K} with RIRs from various datasets \cite{politis:2020:DCASE,orhun_olgun_2019_2635758,mckenzie2021dataset,defferrard2016fma,gotz2021dataset,chesworth2024room,schneiderwind2019data}. We created 5,000 one-minute FOA clips averaging 18 events per clip (std=6), and 500 additional clips for validation. Background noise levels were sampled from a normal distribution (mean=\,–65\,dB, std=15) to encourage robustness. We overrepresented ``Knock'' and ``Bell'' sounds due to their detection difficulty.
Using the stereo SELD data generator provided by the challenge organizers\footnote{github.com/SonyResearch/dcase2025\_stereo\_seld\_data\_generator}, we split each FOA clip into twelve 5-second segments (i.e., with hop size=5s), filtered out silent ones, and applied four random FOA rotations to each segment before extracting the stereo sounds, resulting in $\sim$150k training clips. We'll refer to this synthetic dataset as ``audio 5k'', as it is derived from 5,000 FOA files. 
We employed audio 5k to pre-train the audio backbone of the model, i.e., the SELD encoder and the first CMC.

\begin{figure}[tb]
\begin{minipage}{0.49\columnwidth}
\includegraphics[width=\columnwidth]{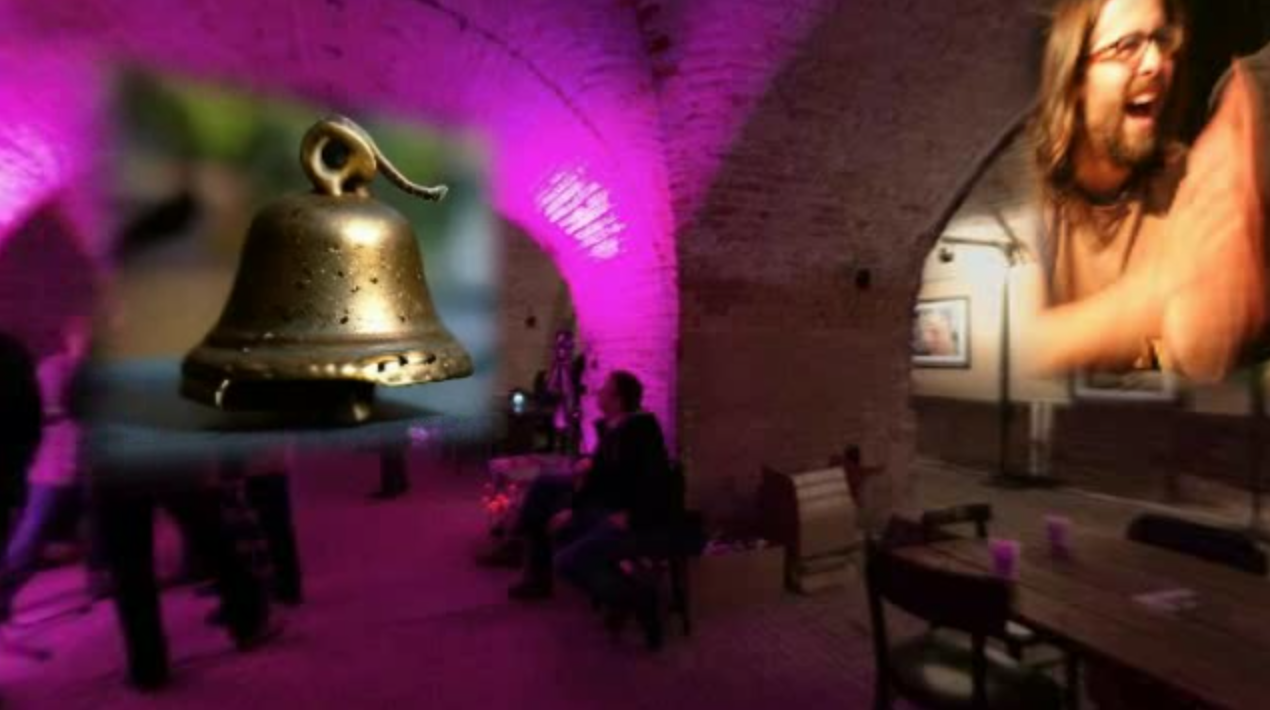}
\vspace{-3mm}
\end{minipage}\hfill
\begin{minipage}{0.49\columnwidth}
\includegraphics[width=\columnwidth]{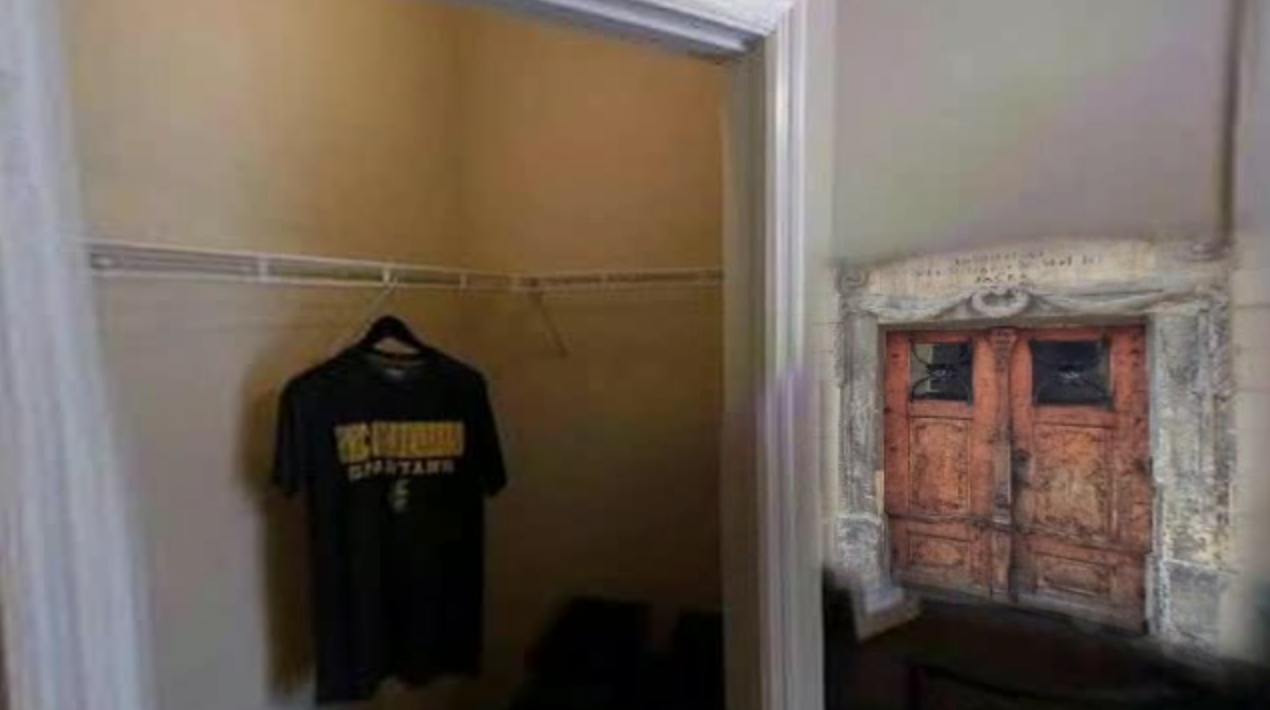} 
\vspace{-3mm}
\end{minipage}

\begin{minipage}{0.49\columnwidth}
\includegraphics[width=\columnwidth]{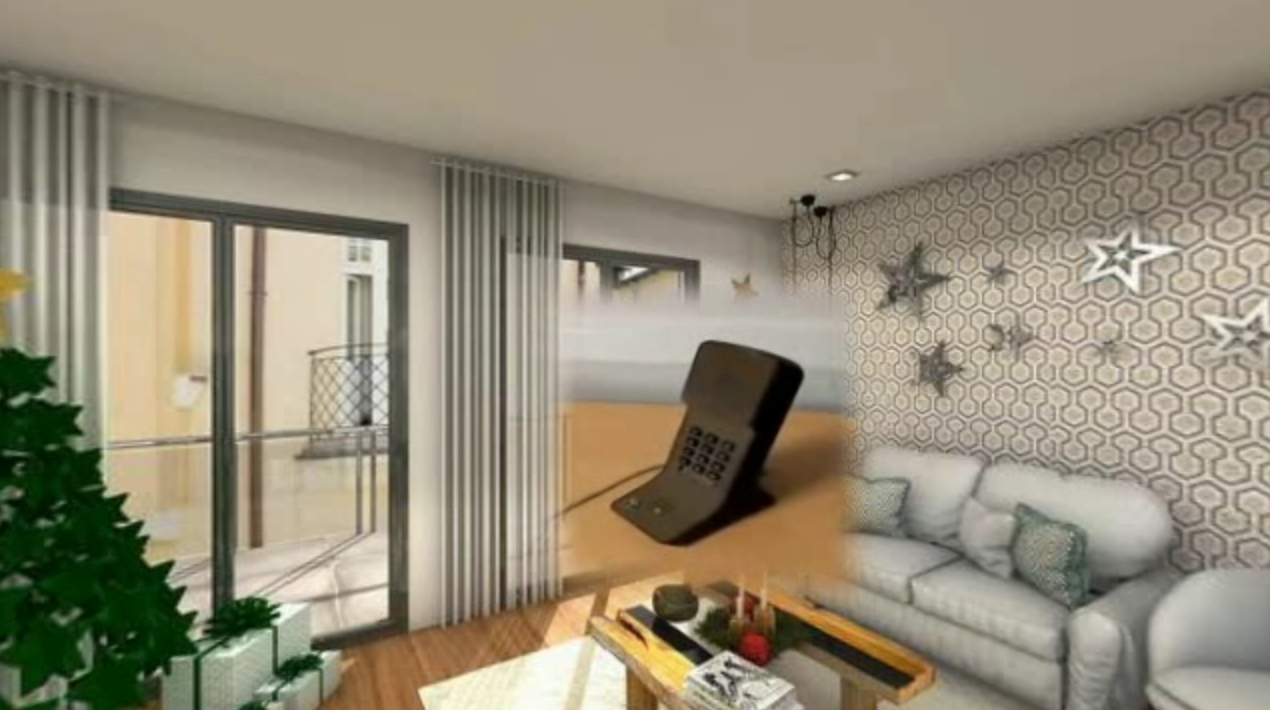}
\end{minipage}\hfill
\begin{minipage}{0.49\columnwidth}
\includegraphics[width=\columnwidth]{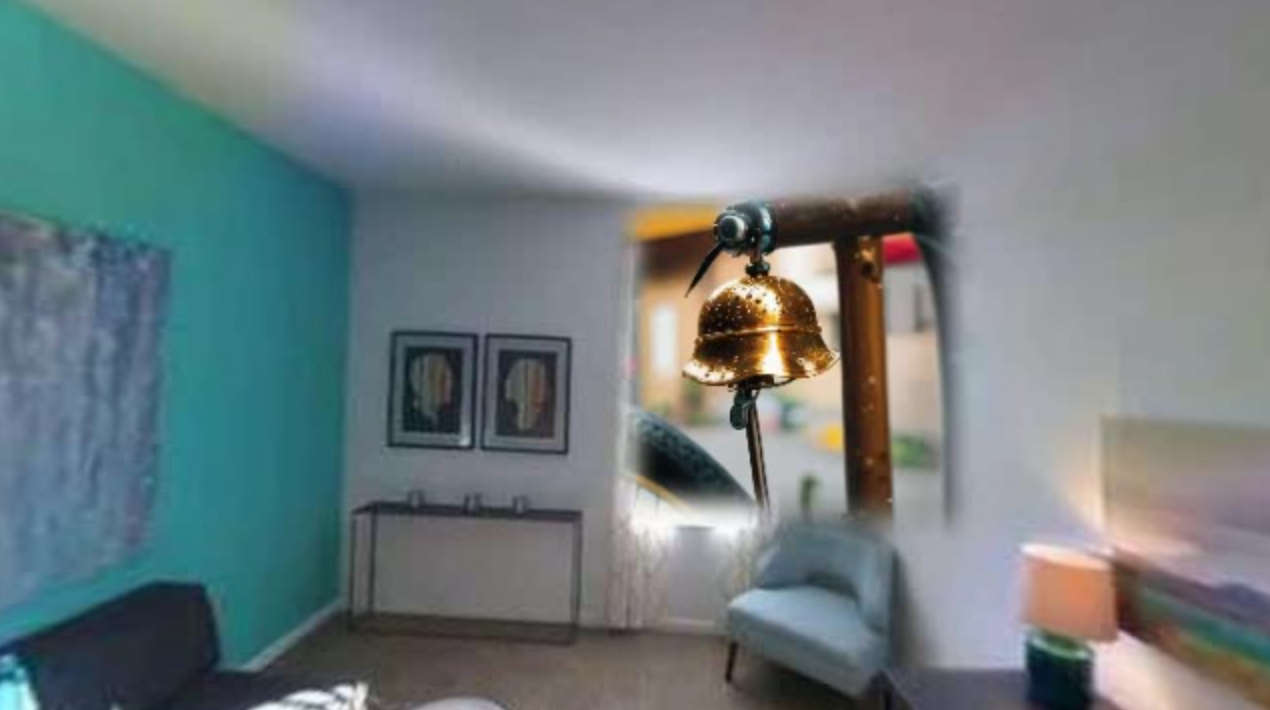} 
\end{minipage}

\captionof{figure}{Examples of synthetic scenes (clockwise from top left) in 
a restaurant, a walk-in wardrobe, a hotel room, a first-floor flat.
The foreground tiles (bell*, laughter, door*, bell*, telephone) are applied with a soft cross-fade to avoid strong edges. *~Generated with NitroFusion \cite{chen:2024:nitrofusion}.}
\label{fig:synthFrames}
\end{figure}

To create an audio-visual synthetic dataset, we generated an additional 2,000 FOA clips with SpatialScaper and synthesized corresponding videos using SELDVisualSynth \cite{roman2025generating}.
SELDVisualSynth creates synthetic videos by placing class-relevant source images or videos (e.g., a person speaking for speech sounds, a door for knock or door sounds) as tiles onto 360$^{\circ}$ background images, based on the positional metadata of sound sources generated by SpatialScaper. 
To enrich the quality of the visual output, we manually curated and extracted class-relevant source images from Flickr30k \cite{Plummer:2015:flickr}. We also selected images of doors from DoorDetect \cite{Arduengo_2021}, which have been employed for ``door open/close'' and ``knock'' sounds. 
Furthermore, we created additional synthetic foreground images for each sound class using NitroFusion \cite{chen:2024:nitrofusion}.  
Instead of employing the background canvases provided with SELDVisualSynth which consists primarily of outdoor environments, we adopted the +2,000 environments of the 360-Indoor dataset \cite{Chou:2020:360indoor}.
We also implemented a soft cross-fade between background canvases and foreground object tiles to remove strong artificial edges from output video.
Fig.~\ref{fig:synthFrames} shows example frames from the dataset.
We initialized the SELD encoder and audio CMC with weights from the audio-only pre-training on audio 5k, and continued training the full model on the +60,000 audio-visual clips generated following this approach. We refer to this dataset here as ``audio-visual 2k''.

To further increase the size of the training dataset and model robustness, we adapted the audio-channel swap (ACS) \cite{Wang:2023:ACS} and video pixel swap (VPS) \cite{Berghi:2024:ICASSP24,Jiang:2024:AVseld,Roman:2024:ehnancedAVseld} augmentations to the stereo scenario. Thus, we swapped the left and right audio channels, and flipped the corresponding video frames to double the size of the training data.
The synthetic data augmented with such methods resulted in over 410h of stereo audio data and an additional 168h of audio-visual data. 
The model was then fine-tuned on the real content of the DCASE2025 Task3 Stereo SELD Dataset.

\section{Experiments}
\label{sec:exp}

\subsection{Implementation Details}


Audio spectrograms were computed using STFT with a 512-point Hann window and 150-sample hop size. At 24\,kHz, this produces 800 temporal bins for the 5-second input clips. For stpACC features, we applied an STFT with a 1014-point Hann window.
This ensures that the autocorrelation covers delays up to approximately 20\,ms after the direct sound. The lag dimension was then downsampled by a factor of 8 to achieve 64 bins and allow concatenation with the other features. More details about stpACC can be found in \cite{Berghi:2025:distanceFeat}.
The SELD encoder and first CMC were pre-trained for 100 epochs on the synthetic audio 5k dataset. The full model was then trained for 80 epochs on the 2k audio-visual set, and fine-tuned for another 80 epochs on the real development set, selecting the best model via F1 score on the test split.
Training used Adam optimizer, batch size 32, and a learning rate reduced by 5\% per epoch after the first 30.

\subsection{Experimental Settings}

Before presenting the results of the proposed training approach, we first introduce an ablation study that isolates the architectural contributions of our model under conditions similar to those used by the challenge baseline systems. We then report a series of experiments based on the full method described in this paper, including additional engineering refinements aimed at further enhancing performance. This reflects our submission to the DCASE 2025 Task 3 Challenge. Our evaluation uses the official metrics of the challenge\footnote{dcase.community/challenge2025/task-stereo-sound-event-localization-and-detection-in-regular-video-content\#evaluation}.

\subsubsection{Ablation Study}

The ablation study is conducted under conditions similar to those adopted by the challenge organizers for the baseline systems. Therefore, we use only left and right log mel spectrograms as audio features (excluding ILD and stpACC) and skip the pre-training stages, training instead on a mix of real clips from the DCASE 2025 Task 3 dataset and 15,000 synthetic audio-visual clips, as in the baseline setup. The ablation includes: a model with only the SELD encoder (SE), one with added CLAP embeddings (SE+CLAP), and the full audio-visual model (SE+CLAP+OV). To match the audio-visual model’s depth, both CMCs in SE and the second CMC in SE+CLAP are retained by setting the input modality Alpha=Beta. We also include an audio-only experiment with left-right channel swap (SE+CLAP+LRS) to assess the impact of this augmentation.

\subsubsection{Proposed Method Evaluation}

%
After the ablation study, we present the results achieved with the full proposed approach.
Since the stereo viewing angle is randomly sampled from the 360$^{\circ}$ frames of STARSS23 \cite{Shimada2023STARSS23AA}, off-screen events outnumber on-screen ones. As a result, the model tends to overpredict the ``off-screen'' class, reaching a stable but biased $\sim$80\% accuracy. To mitigate this, we train a second model with a modified loss function, multiplying the binary cross-entropy loss by 4.0 whenever the ground truth label is on-screen.


Following Jiang \etal \cite{Jiang:2024:AVseld}, we implement a visual post-processing step based on human keypoint detection to refine on/off-screen predictions. Using YOLOv11-Pose \cite{Jocher:2023:yolo11pose}, we extract keypoints and associate specific sound classes with relevant ones: speech and laughter with the nose, clapping with wrist centers, and footsteps with ankle centers. If the predicted DOA falls within 20$^{\circ}$ of the corresponding keypoint direction, the event is marked as on-screen. This step affects only on/off-screen accuracy and $\mathrm{F_{\leq 20^{\circ}/1/on}}$ metrics.
To further improve performance, we apply a majority-vote ensemble across multiple systems. A sound event is considered active at a given time frame if at least two systems detect it and their DOA predictions are within 20$^{\circ}$. The ensemble DOA is then computed as the average of those systems' DOAs. Two exceptions apply: for the ``Bell'' and ``Knock'' classes, a single-system detection is sufficient, following \cite{Berghi:2024:DCASE24techRep}. Similarly, for on/off-screen classification, if any system predicts an event as on-screen, the ensemble output is also on-screen.

\subsection{Results}
\label{subsec:results}

\subsubsection{Ablation Study}

The ablation study results are shown in Table\,\ref{tab:ablation}. The SELD encoder (SE) alone outperforms the baselines by a wide margin, indicating that the CNN-Conformer backbone is highly effective for SELD. Adding the audio CLAP embedding improves the F1 score by nearly one percentage point, as expected given CLAP’s role in enhancing the model's semantic reasoning. Incorporating OWL-ViT, which involves the additional challenge of on/off-screen classification, leads to a further 0.6-point gain in F1 score, although spatial accuracy remains comparable to the SE+CLAP architecture. Finally, the experiment with left-right channel swap demonstrates the benefit of this data augmentation technique.

\begin{table}[bt]
\caption{Ablation results achieved using only left and right log mel spectrograms audio features, and 15,000 synthetic sample during training. $\mathrm{F_{\leq 20^{\circ}/1}}$ (F1) [\%], $\mathrm{F_{\leq 20^{\circ}/1/on}}$ (F1o) [\%], Direction Of Arrival Error (DOAE) [°], On/Off Accuracy (Acc) [\%]. 
AO: audio-only; AV: audio-visual; SE: SELD encoder; CLAP: CLAP audio encoder; OV: OWL-ViT; LRS: left-right swap augmentation.
}
\centerline{\begin{tabular}{l|c|c|c|c|c} \hline
\textbf{Model} & F1\,$\uparrow$ & F1o\,$\uparrow$ & DOAE\,$\downarrow$ & RDE\,$\downarrow$ & Acc\,$\uparrow$ \\ \hline
Baseline AO 
& 22.8 & - & 24.5 & 41.0 & -  \\
Baseline AV 
& 26.8 & 20.0 & 23.8 & 40.0 & \textbf{80.0}  \\ \hline
SE & 34.6 & - & 19.7 & 34.8 & - \\
SE\,+\,CLAP & 35.5 & - & \textbf{19.3} & 34.7 & - \\
SE\,+\,CLAP\,+\,OV & \textbf{36.1} & \textbf{26.3} & \textbf{19.3} & \textbf{32.1} & 79.5 \\ 
\hline \hline
SE\,+\,CLAP\,+\,LRS & \textbf{39.0} & - & \textbf{16.5} & 33.7 & - \\ \hline
\end{tabular}
}
\label{tab:ablation}
\end{table}

\begin{table}[bt]
\caption{Results of the proposed method on the development set DCASE2025 Task3 Stereo SELD Dataset. 
The models audio-only (AO) or audio-visual (AV); variants with weighted loss (+W), visual post-processing (+P); Ensemble (Ens.)\ combines (1),(2),(3) and (4).
}
\centerline{\begin{tabular}{l|c|c|c|c|c} \hline
\textbf{Model} & F1\,$\uparrow$ & F1o\,$\uparrow$ & DOAE\,$\downarrow$ & RDE\,$\downarrow$ & Acc\,$\uparrow$ \\ \hline
(1) AO 
& 45.7 & - & 15.0 & 31.0 & - \\
(2) AO\,/w\,norm.
& 46.0 & - & 15.2 & 30.8 & - \\ \hline
(3) AV 
& 44.4 & 34.0 & 15.6 & 30.4 & 80.5  \\
(3.1) AV+P 
& 44.4 & 34.4 & 15.6 & 30.4 & 80.5  \\
(4) AV+W 
& 45.5& 35.4 & 15.2 & 32.2 & 80.8  \\
(4.1) AV+W+P 
& 45.5 & 35.7 & 15.2 & 32.2 & \textbf{81.0}  \\
(5) Ens. 
& \textbf{48.0} & 37.3 & \textbf{14.0} & \textbf{29.3} & 80.8  \\
(5.1) Ens.+P 
& \textbf{48.0} & \textbf{37.5} & \textbf{14.0} & \textbf{29.3} & 80.8  \\ \hline
\end{tabular}
}
\label{tab:results}
\vspace{-2mm}
\end{table}

\subsubsection{Proposed Method Results}

Table\,\ref{tab:results} shows the results of the full proposed audio-visual method (AV), including two audio-only (AO) experiments based on the SE+CLAP architecture. The two AO variants are identical except for input normalization: the first (system (1)) is fine-tuned using statistics from the development set, while the second (system (2) marked ``/w norm.'') uses statistics from the synthetic 5k dataset used during pre-training. While both perform similarly on the development set, a notable gap emerges from the challenge results on the evaluation subset: AO /w norm. achieves a 42.5\% F1 score, compared to 29.4\% for AO. This suggests that using the same normalization as in pre-training improves generalization, whereas fine-tuning on development set statistics may lead to overfitting.
System (4) uses the same AV model as (3), but with the loss function that weights on/off-screen predictions more heavily when the event is on-screen.
The ensemble model (5) combines predictions from systems (1), (2), (3) and (4).
Finally, applying visual post-processing to (3), (4) and (5) results in the enhanced versions (3.1), (4.1) and (5.1), respectively.

All the experiments substantially outperform the challenge baselines and the other ablation experiments, marking the importance of extensive pre-training with adequate audio input features.
The AO systems achieved slightly better $\mathrm{F_{\leq 20^{\circ}/1}}$ and DOAE scores compared to the base AV systems (3) and (4), which also need to predict on/off-screen labels, adding an extra layer of complexity.
Using a weighted loss function for on/off-screen classification gave only a minor gain in overall on/off accuracy. 
However, it led to a 1.4 percentage point gain in $\mathrm{F_{\leq 20^{\circ}/1/on}}$\@.
The visual post-processing step, which only modifies on-screen predictions, brought small gains in $\mathrm{F_{\leq 20^{\circ}/1/on}}$\@. 
$\mathrm{F_{\leq 20^{\circ}/1}}$, DOAE and RDE remained unchanged as post-processing has no effect on localization and class predictions.
Generally, we observed RDE improvements of about 10 percentage points compared to the baselines. 
These gains may be attributed in part to the inclusion of stpACC features, which enhance the model’s ability to estimate distance.
Model ensembling further improved performance, with system (5) ranking second in the DCASE2025 Task 3 Challenge (Track B).

\section{Conclusion}
\label{sec:concl}

In this work, we address the task of stereo 3D SELD in videos. Our approach integrates semantically rich feature embeddings from CLAP and OWL-ViT within an adapted Conformer module. An ablation study highlights the contribution of each model component, benchmarked against the DCASE2025 Task 3 Challenge baselines. Pre-training on large, curated synthetic audio and audio-visual datasets results in large performance gains.
Further improvements may be sought through stronger integration of auditory and visual modalities exploiting their shared semantics, as well as engineering refinements to model ensembling and post-processing.


\clearpage
\bibliographystyle{IEEEtran}
\bibliography{refs}







\end{document}